*Letter to the Editor*

# Identification of the soft X-ray excess in Cygnus X-1 with disc emission

**M. Bałucińska-Church**[1]**, T. Belloni**[2]**, M. J. Church**[1]**, and G. Hasinger**[3]

[1] School of Physics and Space Research, University of Birmingham, Edgbaston, Birmingham B15 2TT, UK
[2] Astronomical Institute "Anton Pannekoek", University of Amsterdam and Centre for High-Energy Astrophysics, Kruislaan 403, 1098 SJ Amsterdam, The Netherlands
[3] Astrophysikalisches Institut, an der Sternwarte 16, D-14482 Potsdam, Germany

**Abstract.** We present results of a study of the soft X-ray excess in the black hole candidate Cygnus X-1 made with the *Rosat GSPC*, using observations taken during persistent emission at orbital phases close to 0.5. The soft excess can be well fitted as a blackbody with temperature $kT_{bb} = 0.13 \pm 0.02$ keV. $kT_{bb}$ did not vary appreciably with intensity of the source. By assuming that the distance of the source is its lower limit of 2.5 kpc, a luminosity of the soft excess of $4.7 \cdot 10^{36}$ erg s$^{-1}$ was obtained. From this, disc temperatures were calculated as a function of radius, assuming the compact object to be a 10 M$_\odot$ black hole, in particular, the temperature at 7 Schwarzschild radii expected to be highly representative of the total disc emission. This was found to be 0.13 keV, in very good agreement with the spectral fitting result. This good agreement strongly supports the identification of the soft excess with emission from the disc around a black hole.

## 1. Introduction

Cygnus X-1 is a high mass X-ray binary consisting of a O9.7 supergiant and a compact object which is one of the best candidates for a black hole, based on determinations of the mass function and lack of X-ray eclipses (Paczyński 1974). Two periodicities are seen in the data: the 5.6 d orbital period known from optical spectroscopy (Gies & Bolton 1982) and a 294 d periodicity in the X-ray light curve thought to be related to precession of the accretion disc (Priedhorsky et al. 1983). Dips in the X-ray intensity have been observed in the source at orbital

*Send offprint requests to*: M. Bałucińska-Church

phases close to zero, thought to be due to absorption in inhomogeneities in the stellar wind from the companion (Kitamoto et al. 1989; Bałucińska & Hasinger 1991).

The X-ray spectrum is complex, consisting of a number of components. The high energy spectrum shows a break at ∼ 200 keV and has been described by Comptonisation models (eg McConnell et al. 1994). A strong spectral steepening below 10 keV corresponding to transitions to the high X-ray state of the source has been found (Tananbaum et al. 1972).

The presence of a reflection component was identified in *Exosat GSPC* data by Done et al. (1992) who found that an absorption edge at ∼ 6 keV and a broad bump in the spectrum peaking at ∼ 20 keV could be well fitted as a reflection term similar to those seen in AGN. This is a very broad spectral term which contributes to the continuum between 0.1 and ∼ 100 keV. Additionally there is an iron emission line at ∼ 6.2 keV (Barr et al. 1985).

At lower energies the observed spectrum has a soft X-ray excess which became apparent through inconsistencies between the column density $N_H$ obtained from the low energy cut-off of the X-ray spectra and the galactic $N_H \sim 6 \pm 2 \cdot 10^{21}$ H atom cm$^{-2}$ derived from the reddening towards the companion HDE 226868. These inconsistencies were removed if the X-ray spectrum is modelled by a two-component model consisting of a blackbody to represent the soft excess plus a power law to represent the Comptonisation term in the band 1 - 20 keV. The soft excess was first detected by Priedhorsky et al. (1979). Based on the *Einstein/MPC* data, Bałucińska and Chlebowski (1988) showed that this soft excess may contribute up to 5% of the flux in the 1.2 - 20 keV band. A steep low energy spectrum and an emission feature at 0.73 keV were found in *Exosat* transmission grating spectrometer (TGS) data in the band 0.4 - 2.0 keV by Barr and van der Woerd (1990). The presence of the soft excess was confirmed by extensive studies of the *Exosat ME* observations by Bałucińska and Hasinger (1991), who showed that the excess could be modelled as a blackbody with temperature $kT_{bb} \sim 0.25$ keV. However, the ME data were not able to constrain the temperature very well, as the peak of the blackbody spectrum was outside the range of the instrument.

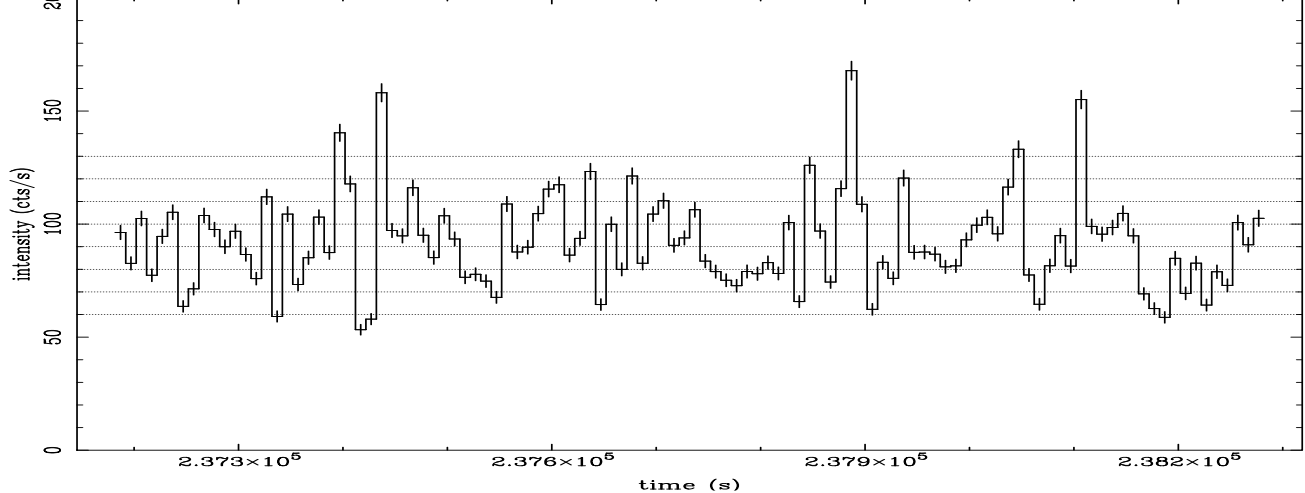

**Fig. 1.** X-ray light curve for the Rosat observation at phase 0.46 in the energy band 0.1 - 2.0 keV with 10 s timebins. The dotted horizontal lines correspond to the intensity ranges in which spectra were accumulated

The *Rosat PSPC* allows a proper study of the soft excess in the appropriate energy band containing the peak of the emission. We present detailed spectral fitting analysis of the soft excess in the band 0.1 - 2.0 keV, from observations made when Cygnus X-1 was in its low state. Results are given here for the part of the observation at orbital phase close to 0.5 where there are no dips and at which the effect of extra absorption due to the stellar wind of the companion is expected to be minimal. Results from the rest of the observation in which intensity dips are seen will be given in a second paper.

## 2. Data analysis

The observations were made on 1991 April 18 - 20 spanning a total of 66 h and consisting of three sections of continuous observation lasting 2030 s, 1890 s and 1100 s. The first two sections were close to orbital phase zero and we do not discuss this data here. The third section was at orbital phase 0.46 calculated using the accurate ephemeris of Gies and Bolton (1982). Source data were extracted from a circle of radius $0.21°$ centred on the source to include photons scattered outwards in the dust halo (Predehl & Schmitt 1995). Background was extracted from an annulus extending from $0.22°$ to $0.29°$. The background was subtracted and the data were corrected for deadtime and vignetting. We present results below obtained using the March 1992 response function appropriate to this observation in the early part of the *Rosat* mission (see Fiore et al. 1994). However in order to assess systematic errors, we also analysed the data using the January 1993 response, and found that parameters derived from spectral fitting changed by only 1.5% to 3%, giving confidence in the results.

The light curve of the third section of data is shown in Fig. 1. It can be seen that, because of the strong variability of the source on short timescales, it is not possible to accumulate data for a sufficiently long time with the source at a constant intensity level to produce a spectrum of suffficient quality (requiring $\sim 100$ s of data). Consequently the data were divided into seven intensity bins, each 10 c/s wide covering the ranges from 60 - 70 c/s to 120 - 130 c/s in the band 0.1 - 2.0 keV. These intensity bins were equivalent typically to 130 s of data, so that Poisson errors were sufficiently small for good fitting to be made. For each of these intensity bands, a spectrum was produced by rebinning into 24 channels. We added conservatively a systematic error of 2% to each channel. If the actual error is somewhat smaller, the effect will be that the errors we give for parameters derived from spectral fitting should be reduced. The values of reduced $\chi^2$ would have to be increased by 10 - 20%, so that unacceptable models become more unacceptable; however for our best fit models $\chi_r^2$ would remain less than unity.

Additionally, a total spectrum was produced including all intensities with 24 channels to allow searching for emission and absorption features.

## 3. Results

Firstly, spectral fits of the spectra in the 7 intensity bins were tried using simple models including an absorbed power law model. In this case, the quality of the fits was unacceptable, with reduced $\chi^2$ typically as large as 1.24 with 21 degrees of freedom. Moreover, the power law photon index $\Gamma$ averaged over the 7 intensity bands was $2.3 \pm 0.2$ ($1\sigma$), and the individual values of $\Gamma$ in all 7 bands were significantly higher than the widest range of possible values we obtained in the higher *Exosat ME* energy band, between 1.5 and 1.8 (Bałucińska & Hasinger 1991). This clearly demonstrates the presence of the soft excess. The *Rosat PSPC* is not able to constrain power law the band 1 - 10 keV or higher is needed. It is however very well able to determine the parameters of a thermal component peaking within the PSPC band, and so the most sensible approach is to use the PSPC data in conjunction with $\Gamma$ values previously established for the source. For this we have used results from observations of Cygnus X-1 with the TTM and HEXE instruments onboard *Mir*. These instruments are normally used simultaneously and so provide high quality spectra extending over the very broad 1 - 200 keV band. Consequently continuum parameters may be well established and we have previously found that from several TTM/HEXE observations $\Gamma$ for the intrinsic power law has the value $1.73 \pm 0.02$ (Bałucińska-Church et al. 1996).

Fitting the simple absorbed power law model with $\Gamma$ fixed at 1.73 gave very poor fits in all 7 intensity bands with $\chi_r^2$ typically 1.7. Adding a second soft power law to the model to parameterise the spectra was next tried, and in all cases this required $\Gamma \sim 8$, clearly unphysical. The large inconsistency between this value and 1.73 is further evidence for the soft excess. Thus we were able to discard a simple one-component model, and adopted a physically more realistic two-component model consisting of a power law with $\Gamma$ fixed at 1.73, plus a blackbody to model the soft excess, with a single absorption term for both components. Good fits were obtained in all 7 intensity bands, with reduced $\chi^2$ significantly improved over simple models.

isation clearly increasing linearly with intensity.

The mean value of $kT_{bb}$ for all intensity bands was $0.13 \pm 0.02$ keV, showing that the PSPC was able to locate the blackbody peak and so determine the value of $kT_{bb}$ with good accuracy. It was not necessary to specify a separate column density for each component, and the mean value of $N_H$ was $7.5 \pm 1.0 \cdot 10^{21}$ H atom cm$^{-2}$.

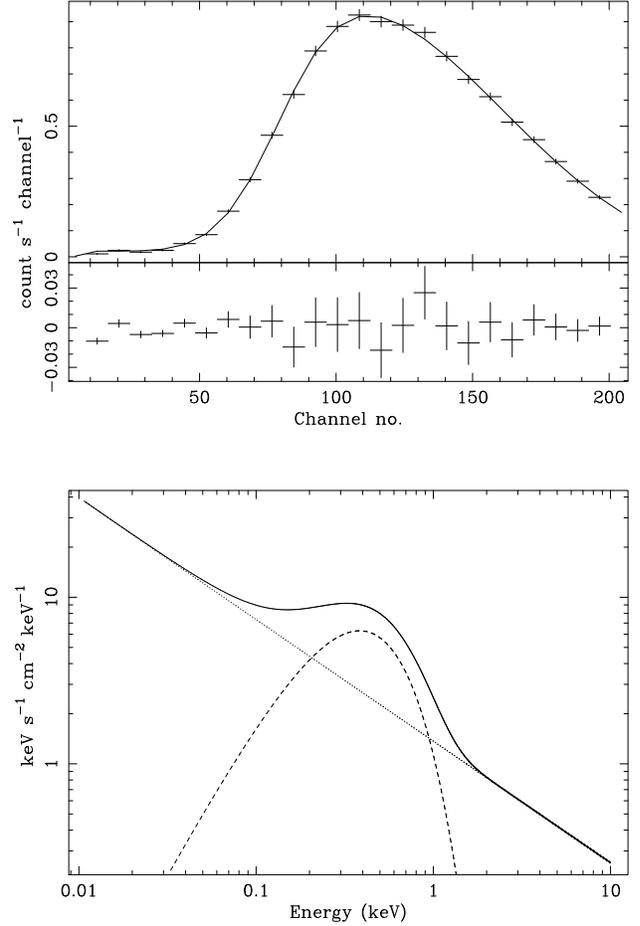

**Fig. 3.** Top panel: Spectral fit to the data in the total spectrum using the two-component model, together with residuals. Bottom panel: individual model component contributions to the unabsorbed total energy spectrum. The dotted and dashed lines show the power law and blackbody components respectively

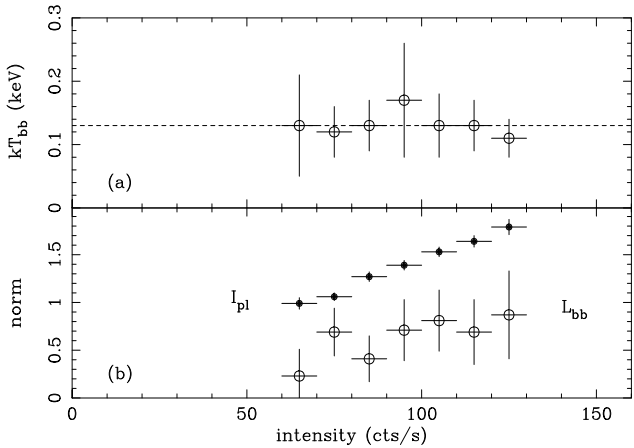

**Fig. 2.** (a) Spectral fitting results for the 7 intensity levels into which the observation was divided showing blackbody temperature $kT_{bb}$ as a function of c/s. The dashed line shows the mean value. (b) Variation of the power law normalisation at 1 keV $I_{pl}$ (photons cm$^{-2}$ s$^{-1}$ keV$^{-1}$) and the blackbody normalisation $L_{bb}$ (expressed as luminosity at 10 kpc in erg s$^{-1}$) with intensity

In Fig. 2a, results for the blackbody temperature $kT_{bb}$ as a function of count rate are shown. It can be seen that there is no variation of $kT_{bb}$ with intensity, showing that the short-term variability apparently takes place without change in the shape of the spectrum. Only the normalisations of the power law and

We tested the robustness of our result for $kT_{bb}$ by fitting the total spectrum, adding all intensity bands, with the same two-component model, and fixing $\Gamma$ at values covering the widest range of acceptable values: 1.5 to 1.8. For these extreme values of $\Gamma$, $kT_{bb}$ changed from $0.15 \pm 0.02$ keV to $0.13 \pm 0.02$ keV; ie not significantly different from the value we previously obtained in the individual intensity bins.

In Fig. 3 we show the best fit to the total spectrum together with residuals with photon index $\Gamma$ fixed at 1.73; the individ-in the lower panel. We see no evidence for emission or absorption features, particularly the feature at $\sim 0.7$ keV reported by Barr and van der Woerd (1990). Best fit values from fitting the total spectrum are: $N_H = 7.0 \pm 1.0 \cdot 10^{21}$ H atom $cm^{-2}$, blackbody bolometric luminosity at 10 kpc: $L = 7.5^{+9.0}_{-5.3} \cdot 10^{37}$ erg s$^{-1}$, $kT_{bb} = 0.14^{+0.02}_{-0.01}$ keV, power law normalisation at 1 keV $= 1.37 \pm 0.12$ photons $cm^{-2}$ s$^{-1}$ keV$^{-1}$ (90% confidence errors). It can be seen that the blackbody luminosity is non-zero at a high confidence level; the value is greater than $2 \cdot 10^{37}$ erg s$^{-1}$ at 99% confidence. The absorbed flux of the blackbody in the band 0.1 - 2.0 keV is $2.3 \cdot 10^{-10}$ erg s$^{-1}$, the power law absorbed flux in the same band is $9.0 \cdot 10^{-10}$ erg s$^{-1}$, so that the blackbody comprises 20% of the energy flux in this band. For unabsorbed spectra, this fraction rises to about 50%.

## 4. Discussion

We have investigated further the lack of any spectral feature at $\sim 0.7$ keV, reported by Barr and van der Woerd (1990). Using their best fit results we have simulated the *Exosat TGS* data, and find that if the feature was present at the same strength in our *Rosat* observation, it would have been easily detectable. However the PSPC data had a 25% higher flux in the band below 0.7 keV, but a 25% smaller flux above 0.7 keV; ie there were significant changes in both the softness and brightness of the source.

The results consistently show a temperature for the soft excess of $\sim 0.13$ keV. We now estimate the temperature of the emission expected from an accretion disc about a 10 M$_\odot$ black hole, and show that there is good agreement.

It is expected that the disc thermal emission will be modified by reflection; however the work of Ross and Fabian (1993) has shown that for reflection in the disc around a 10 M$_\odot$ black hole, and in the restricted band 0.3 to 3 keV (Fig. 5, Ross & Fabian) the incident blackbody is not strongly modified. Furthermore, Ross et al. (1992) showed that the temperature of the accretion disc at 7 Schwarzschild radii, where the contribution to the luminosity $r^2 F$ peaks (F is the emitted flux), is highly representative of the total disc emission from the multi-temperature disc. They also showed that fitting a synthesised disc spectrum with a simple blackbody produces results for $kT_{bb}$ in reasonable agreement with the temperature at $7 r_S$.

Firstly we calculate the luminosity of the soft component at the time of observation, using our best fit spectral parameters and setting the distance of the source at its lower limit of 2.5 kpc (Bregman et al. 1973). This was found to be $L_x = 4.7 \cdot 10^{36}$ erg s$^{-1}$. However the actual distance of the source is poorly known and the luminosity could be several times larger. This luminosity was converted into mass accretion rate $\dot M \sim 5.2 \cdot 10^{16}$ g s$^{-1}$, assuming an efficiency of 0.1, and from this we calculated disc temperatures as a function of radius in a multi-temperature disc (eg Frank, King & Raine 1992). At 7 Schwarzschild radii (210 km), we find a temperature $kT_{bb}$ of $0.13 \pm 0.03$ keV, where the errors reflect only the uncertainties in the soft component luminosity from spectral fitting. Thus, there is good agreement of the observationally determined soft tance of the source is appreciably greater than 2.5 kpc, the luminosity will be higher, however temperatures in the disc are proportional to $\dot M^{1/4}$ and so a factor of 2 in distance implies only a 40% increase in temperature.

The blackbody luminosity is less than 1% of the Edddington luminosity $L_E = 1.47 \cdot 10^{39}$ erg s$^{-1}$ for a 10 M$_\odot$ object with 70% hydrogen. The luminosity of the power law component is estimated as $\sim 1.9 \cdot 10^{37}$ erg s$^{-1}$. Thus the above comparison of the temperature of the soft excess with disc temperatures is for the low state of the source at a low fraction of the Eddington limit.

Using the total spectrum, the emitting area of the blackbody was found to be $1.3 \cdot 10^{16}$ cm$^2$ from the luminosity of this component and the temperature $kT_{bb} = 0.14$ keV. If this is emission from a disc, the outer radius would have to be $\sim 460$ km, so that the emission region effectively extends from the inner disc to $\sim$ 15 Schwarzschild radii.

*Acknowledgements.* Analysis was carried out at the University of Birmingham using the facilities of the PPARC Starlink node. TB is supported by NWO under grant PGS 78-277.